\documentclass[twocolumn, aps,prl]{revtex4-1}
\usepackage{graphicx}
\usepackage{amsfonts}
\usepackage{amsmath}
\usepackage{textcomp}
\usepackage{amssymb}
\usepackage{bm}
\usepackage{color}
\usepackage{subfigure}

\newcommand{\ev}[1]{{\left\langle{#1}\right\rangle}}
\newcommand{\ket}[1]{{\left|{#1}\right\rangle}}

\newcommand{\bra}[1]{{\left\langle{#1}\right|}}
\newcommand{\ip}[2]{{\left\langle{#1|#2}\right\rangle}}

\newcommand{\g}{\mathbf{g}}

\newcommand{\p}{\mathbf{p}}

\newcommand{\rr}{\mathbf{r}}
\newcommand{\dd}[1]{\mathrm{d} #1\,}

\newcommand{\smi}[1]{\sigma_{#1}}

\newcommand{\al}{\boldsymbol{\alpha}}

\begin{document}

\title{Interferometry with Synthetic Gauge Fields}
\author{Brandon M. Anderson}
\author{Jacob M. Taylor}
\author{Victor M. Galitski}
\affiliation{Condensed Matter Theory Center and
Joint Quantum Institute, Department of Physics, University of
Maryland, College Park, MD 20742-4111}

\begin{abstract}
We propose a compact atom interferometry scheme for measuring weak, time-dependent accelerations. Our proposal uses an ensemble of dilute trapped bosons with two internal states that couple to a synthetic gauge field with opposite charges. The trapped gauge field couples spin to momentum to allow time dependent accelerations to be continuously imparted on the internal states. We generalize this system to reduce noise and estimate the sensitivity of such a system to be $S\sim 10^{-7} \frac{\textrm{m} / \textrm{s}^2}{\sqrt{\textrm{Hz}}}$.
\end{abstract}

\maketitle

In recent years atom interferometry has emerged as a powerful tool for
precision gravimetry and accelerometry.~\cite{AGClauser, AGReview1,
  AGReview2} Experiments such as ~\cite{GRKasevich1, AGChu1, AGChu2,
  GRKasevich2, GKasevich1, GKasevich2} are the most accurate
measurements to date of surface gravity, certain fundamental
constants~\cite{NKasevich, AGBiercuk, NBertoldi} and also provide probes of
General Relativity and the inverse square law~\cite{GRKasevich1,
  GRKasevich2, GRChu, GRZoest}.  Furthermore, accelerometers have wide
application in more practical settings such as inertial navigation,
vibration detection, and gravitational anomalies such as oil
fields~\cite{NIS}. Current experiments use short Raman pulses to manipulate spin
states followed by periods of free evolution, corresponding to free
flight, to accumulate sensitivity to external fields. During free
flight, a sensitivity to external fields is imparted on the internal
spin states in the form of a path-dependent phase. This phase can then
be measured through a final Raman pulse and spin-dependent
fluorescence techniques.

%
At the same time that interferometry has emerged as a tool, interest
in synthetic gauge fields has also arisen, mostly in the context of
the quantum Hall Effect~\cite{SMFLith1, SMFQHE} and cold atom spintronics~\cite{SMFSpintronics, SMFNonAbelian, SMFSinova, SMFLith2}. These systems use
optical coupling of internal spin states, simultaneous with momentum
exchange with Raman laser beams, to induce an effective vector
potential. Depending on the optical configuration these setups can
simulate systems such as spin-orbit coupling~\cite{SMFSOBEC, SMFSpielman1},
monopoles~\cite{SMFLith2}, or a constant magnetic field~\cite{SMFSpielman2,
  SMFSpielman3}. 

The optical coupling to the internal degrees of freedom provides a continuous coupling of momentum and spin. This is in contrast to standard interferometry schemes where spin and momentum coupling is generated only through a set of discrete Raman $\pi/2$ and $\pi$ pulses. In this paper we propose a new type of interferometer that uses the spin-momentum coupling to measure AC signals. We use the continuous spin-momentum compuling of the gauge field to produce an interferometer sensitive to high frequency time dependent (or AC) fields. This is in contrast to current systems that at best are sensitive to constant (DC) or low frequency $\leq 1 \textrm{Hz}$ signals~\cite{AGWave1, AGWave2}.  We specifically propose using a trapped system of cold bosons under the influence of an optically induced gauge field to measure weak AC gravity signals. We discuss some potential implementations and we estimate that such a system will have a sensitivity of $S \sim 10^{-7} \frac{\textrm{m}/\textrm{s}^2}{\sqrt{\textrm{Hz}}}$. We note that since our system is trapped it can be implemented on an atom chip~\cite{TAGKetterle1, TAGSchmiedmayer, TAGHansch}.



\begin{figure}[!t]
\centering
\includegraphics[width=\columnwidth]{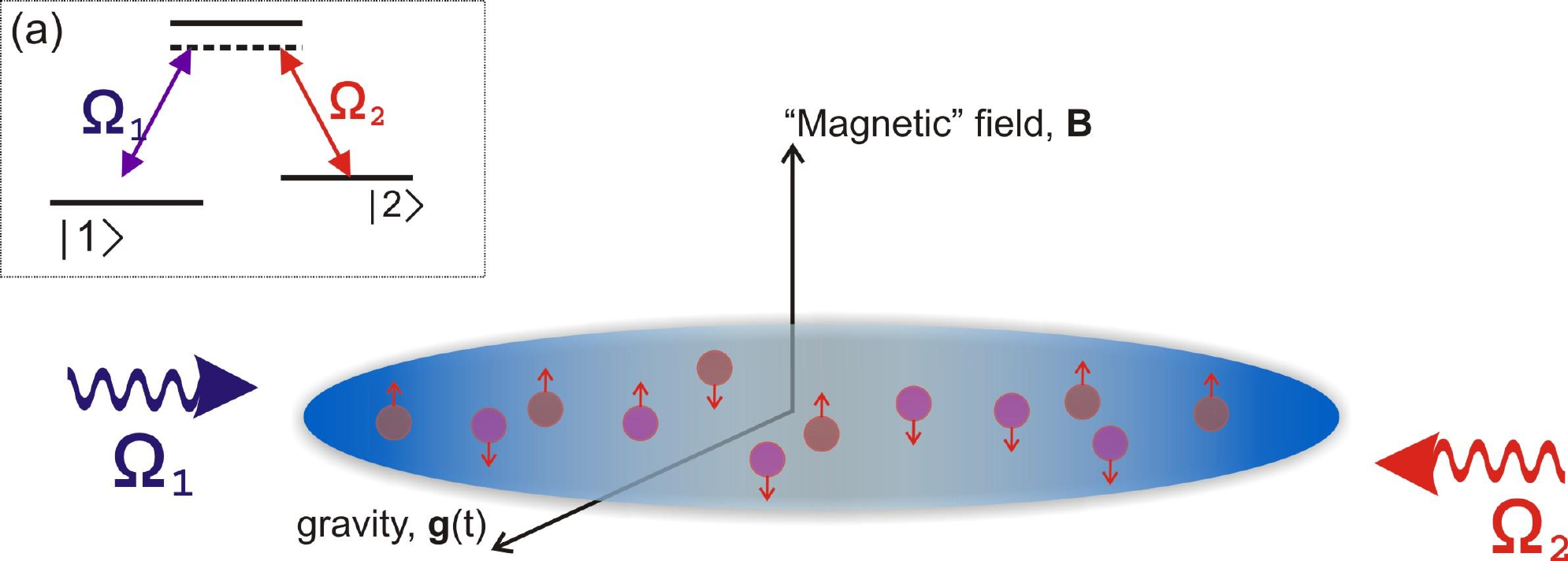}
\caption{\label{fig:setup} A potential implementation of our interferometer based upon Ref.~\cite{SMFQHE}. The Raman beams $\Omega_{1,2}$ couple a three level atom by two parallel Gaussian profiles with peaks that are spatially offset. Two of the dressed states become degenerate in the large detuning limit, $\Delta \rightarrow \infty$, and couple to a ``synthetic gauge field'' with opposite charges.}
\end{figure}

As a toy model, we consider a single particle with an internal degree of freedom (pseudo-spin) in a harmonic trap with spin-orbit coupling and an external force:
\begin{equation}
H = \frac{\left[\hat{\p} - \sigma \mathbf{A}(\hat{\rr})\right]^2}{2m} + \frac{1}{2} m \omega_0^2 \hat{\rr}^2 - m \g(t) \cdot \hat{\rr} \label{eq:H}
\end{equation}
where $\hat{\p}$ and $\hat{\rr}$ are the position and momentum operators respectively, $\omega_0$ is the trapping frequency, $\sigma=\pm 1$ labels the pseudo-spin of the particle, m is the mass of the particle, $\g(t)$ is the time dependent external force and $\mathbf{A}(\hat{\rr})$ is the spin-orbit coupling field, or vector potential. We confine the particle to a two dimensional plane and chose vector potential to have the form of a magnetic field $\mathbf{A}(\hat{\rr}) = m \omega_c x \hat{y}$, where $\omega_c$ is the characteristic frequency scale of the spin-orbit coupling and $\hat{e}_z$ is the unit vector perpendicular to the plane of confinement. This toy model captures the ideal behavior of systems such as Ref.~\cite{SMFSpielman3, SMFQHE}.


Without spin-orbit coupling, $(\omega_c \rightarrow 0)$, the path of the particle will depend on the force $\g(t)$ and will be independent of spin. With spin-orbit fields the path will depend on the spin of the particle as well as the force. In an atom interferometer, path differences can be mapped to an interference signal by creating an initial superposition of two spin states and sending them on spin-dependent trajectories. The phase picked up on a semi-classical path can be found by taking the variation of the action S in the path integral formulation of quantum mechanics. 

For the system in Eq.~\ref{eq:H} we find the first order phase due to $\g$ is
\begin{equation}
e^{i S_{cl}} = e^{i \frac{m}{\hbar} \int \dd{t} \rr(t) \cdot \g(t)} \label{eq:phase}.
\end{equation}
We become sensitive to this phase by creating an initial spin superposition $\frac{\ket{1}+\ket{-1}}{\sqrt{2}} \ket{\phi}_{orbital}$ which evolves to $\frac{1}{\sqrt{2}}\left( e^{i S_1} \ket{1} \ket{\phi_1} + e^{i S_{-1}} \ket{-1} \ket{\phi_{-1}}\right)$. In general the spin states $\ket{\pm 1}$ are entangled to the orbital states $\ket{\phi_{\pm 1}}$. In order to measure the phase we desire a pure spin measurement. We chose the paths to have complete orbital overlap, $\ip{\phi_1}{\phi_{-1}}=1$, at the time of phase measurement. This places the system in the pseudo-spin state $\frac{1}{\sqrt{2}} (\ket{1} + e^{i \Delta S} \ket{-1})$ which allows us to measure $\Delta S = S_{-1} - S_{1}$ though a single operator measurement such as $\hat{S}_y$.

The physics described in our toy model is an idealized version of the proposal given in Ref.~\cite{SMFQHE}, although other setups such as the experiment by Y.J. Lin \textit{et. al}~\cite{SMFSpielman3} have similar physics. This setup uses a cold atom in $\Lambda$-scheme as can be seen in the inset of Fig.~\ref{fig:setup}. The Raman beams used have a Gaussian profile with offset centers which give spatial dependence to the dressed states. This spatial dependence induces dynamics that is identical to a charged particle in a magnetic field. In the large detuning limit, $\Delta \rightarrow \infty$, one of the bright states becomes degenerate with the dark state, however the ``charge'' that these two states see has opposite sign. We note that the synthetic field in this setup is non-uniform. This leads to further technical difficulties but does not change the underlying physics of our proposal.

We now detail the specific solution for our spin-orbit coupled system described by Eq. \ref{eq:H}. We start by solving the Heisenberg equations of motion for the system. These correspond to the classical, spin-dependent, Hamilton equations of motion since the system is quadratic in $\rr$ and $\p$. Not surprisingly, these solutions correspond to a combination of cyclotron orbits and orbits around the trap center. The direction of the orbits is set by the pseudo-spin, $\sigma$.

\begin{figure}[!t]
\centering
\subfigure[]{\includegraphics[width=.4\columnwidth]{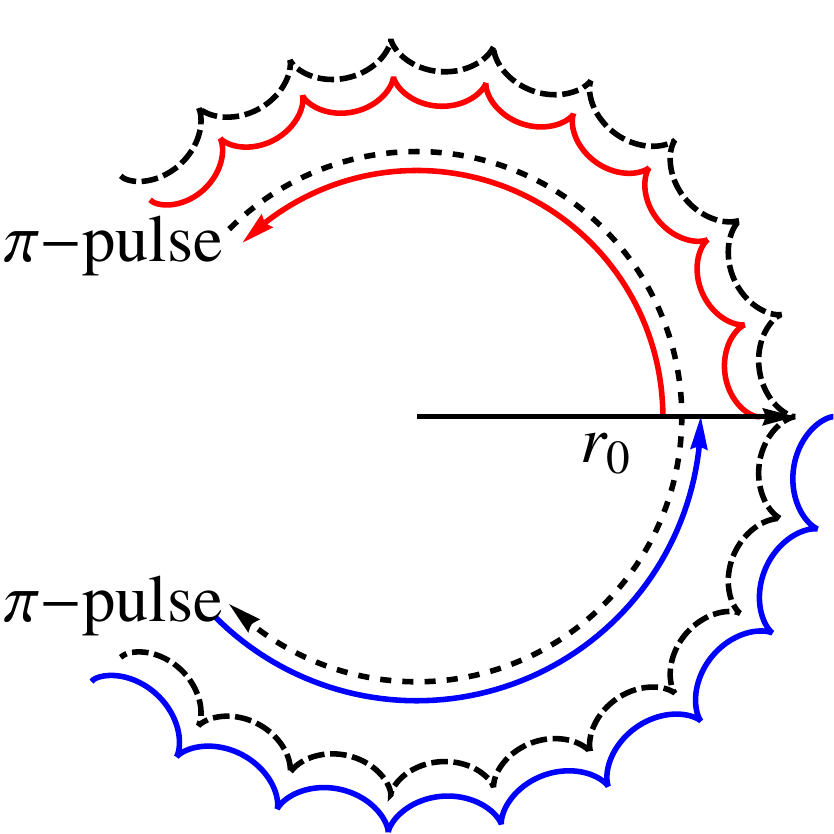}\label{fig:cppulse}}
\hspace{.05\columnwidth}
\subfigure[]{\includegraphics[width=.4\columnwidth]{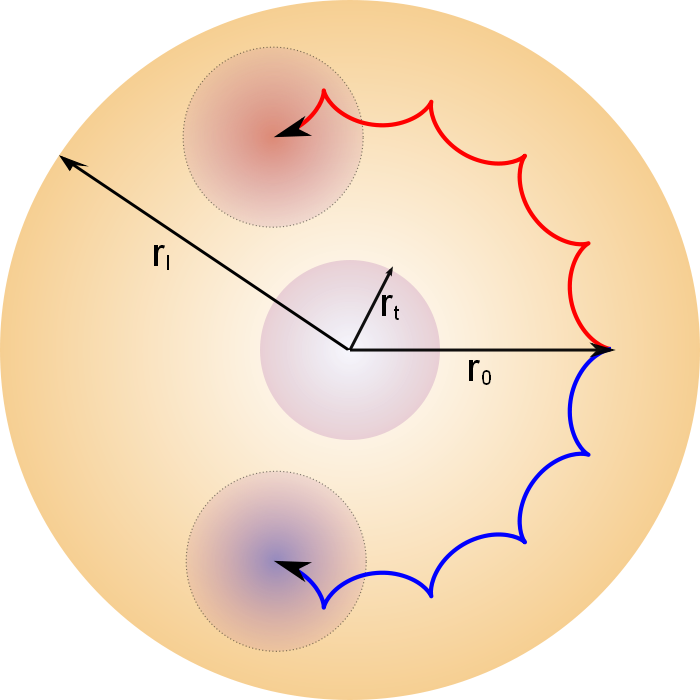}\label{fig:experiment}}
\subfigure[]{\includegraphics[width=.49\columnwidth]{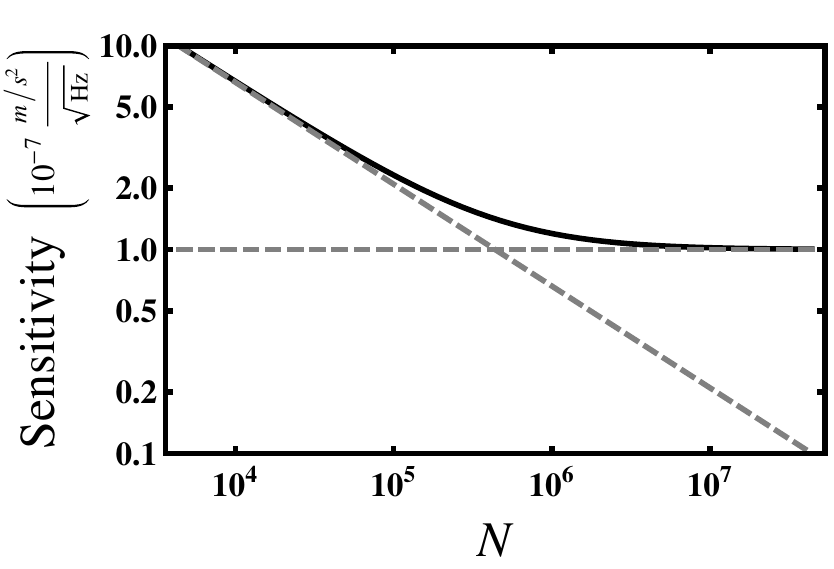}\label{fig:sens}}
\subfigure[]{\includegraphics[width=.49\columnwidth]{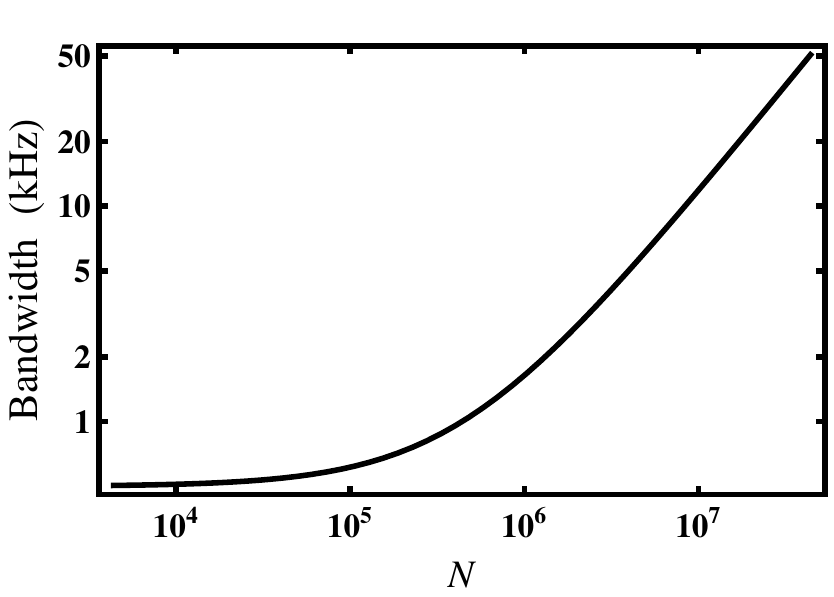}\label{fig:bw}}
\caption{ \ref{fig:cppulse}The classical path a particle will follow with the CP pulse sequence given by Eq.~\ref{eq:Ucp}. The red path corresponds to the initial free evolution for a time $t$. The dashed path is the time reversed path allowed to evolve for a time $2 t$. Finally, the solid blue path is the return trajectory for another $t$. Note that the three trajectories will overlap in practice, and have been offset just for a visual aid. The three arcs correspond to the direction of motion of the classical trajectory. We have also chosen $\epsilon = \omega_+/\omega_- = 22$. \ref{fig:experiment} A schematic of the experimental setup. A thermal cloud of radius $r_t$ is displaced by $r_0 = r_l-r_t$, which is limited by the laser inhomogeneity radius, $r_l$. The two spin states orbit along trajectories mirrored across $\rr_0$. \ref{fig:sens} The dependence of the sensitivity of the system based upon atom number. Below $N_c \sim 10^6$ the sensitivity grows as $\frac{1}{\sqrt{N/N_c}}$. Above $N_c$ the sensitivity  $S \sim 10^{-7} \textrm{m}/\textrm{s}^2 / \sqrt{\textrm{Hz}}$ is independent of the number of particles. \ref{fig:bw} The bandwidth of the system with optimal sensitivity as a function of the number of particles. }
\end{figure}

The solutions are characterized by the two frequencies $\omega_\pm = \tilde{\omega} \pm \omega_c /2$, with $\tilde{\omega}^2 = \omega_0^2 + (\omega_c/2)^2$. We will consider initial conditions given by $\rr(0) = \rr_0$ and $\dot{\rr}(t) = 0$, as will be discussed later. Fig. \ref{fig:experiment} shows these paths in the absence of a driving field. The sign of the charge changes the direction of cyclotron motion so the paths are mirrored along $\rr_0$. We expect a driving force to break the mirror symmetry of the paths. 

In a quantum system this symmetry breaking perturbation will result in a pseudo-spin dependent phase. We can exploit this phase to make an interference measurement.  Specifically, we
would prepare the system in an initial state $\ket{0,\uparrow}$, where
$\ket{0}$ is the orbital ground state of the system and $\smi{z}
\ket{\uparrow} = \ket{\uparrow}$. Next we place the system in a superposition of pseudo-spin states with a $R_{\hat{y}}(\pi/2) = e^{-i \sigma_y \pi / 4}$ Raman pulse.  Then, we suddenly
displace the minimum of the harmonic trap by an amount $\rr_0$.  If we
allow the system to evolve freely in time the two different spin
states will follow time reversed classical trajectories. In the process, they also accumulate a pseudo-spin-dependent
phase term $e^{i \smi{z} (\hat{z} \times \rr_0) \cdot \int \dd{t}
  \g(t) \, h_\bot(t)}$, where $h_\bot(t) = \frac{1}{2
 \tilde{\omega}} \left( \omega_- \sin(\omega_+ t) - \omega_+  \sin(\omega_- t) \right)$. We now wait for a time $t$ at which
the two classical trajectories overlap again. We then use a
$R_{\hat{y}}(-\pi/2)$ pulse to convert the coherence into a population. 
A $S_z$ spin measurement will give
\begin{equation}
\ev{S_z} =\sin\left[2 \int_0^{t^\prime} \dd{t} (\hat{z} \times \rr_0) \cdot \g(t^\prime) \, h_\bot(t^\prime) \right]. \label{eq:Sz}
\end{equation}
where we have neglected higher order terms in $g / l_o \omega^2$ with the harmonic oscillator length $l_o = \sqrt{\hbar / m \tilde{\omega}}$. We can represent the above pulse sequence as the unitary matrix $U_p = R_{\hat{y}}(-\pi/2) U(t) D[R \rr_0] R_{\hat{y}}(\pi/2)$ where $D[\rr_0]$ is a spatial displacement of $\rr_0$ and  the time evolution operator $U(t)$ can be found exactly. The expectation value in Eq. \ref{eq:Sz} is then given by $\ev{S_z} = \bra{0,\uparrow} U^\dagger_p S_z U_p \ket{0,\uparrow}$ and can be shown to reproduce Eq.~\ref{eq:Sz}.

\begin{figure}[!t]
\centering
\includegraphics[width=\columnwidth]{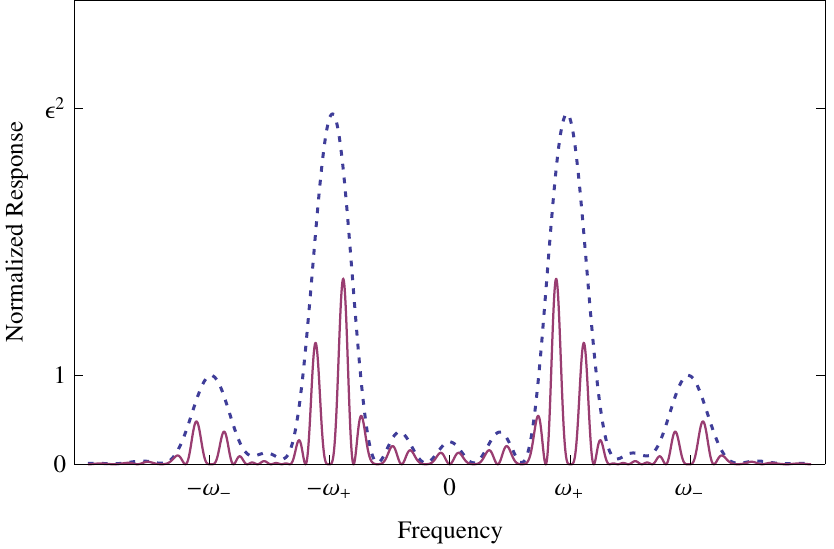}
\caption{\label{fig:response} The normalized response function for $\left|\frac{\tilde{\omega}  F(\omega)}{r_0 \Delta t}\right|^2$ for the pulse sequence $U_p$, (dashed) or $U_{C-P}$, the Carr-Purcell like pulse sequence (solid). For both sequences we used $\Delta t = \frac{10 \pi}{\omega_+ + \omega_-} = \frac{5 \pi}{\tilde{\omega}}$. The frequency of the zeroes between the peaks are given by $\omega = \omega_\pm$, and the relative peak amplitudes are given by $\epsilon = \omega_+ / \omega_-$. Note we have scaled the response for to the $CP$ pulse sequence by a factor of 16 to account for the factor of four increase in interrogation time.}
\end{figure}

We now find the response of our interferometer to an arbitrary time varying force. We can express our spin population measurement as $\ev{S_z} = \sin\left[ \int \frac{\dd{\omega}}{2\pi} \tilde{g}_\bot(\omega) F_0(\omega) \right]$ where $\g(t) = \int \frac{\dd{\omega}}{2\pi} e^{-i \omega t} \tilde{\g}(\omega)$ and 
\begin{equation}
F_0(\omega) = \frac{i r_0 t}{\tilde{\omega}} \sum_{\{\sigma,\tau = \pm 1\}} \sigma \tau \, \omega_{{-}\sigma} \, f(\omega + \tau \omega_\sigma) \label{eq:f0}
\end{equation}
is the response function of the system, where $f(\omega) = \frac{\sin(\omega t /2)}{\omega t /2} e^{-i \omega t /2}$.
The behavior of the response function can be seen in
Fig. \ref{fig:response}. The peak response of the system is at the frequencies $\omega = \omega_\pm$ with relative peak amplitudes of
$\omega_-/\omega_+$. The bandwidth of the system varies with $1/t$ giving a large bandwidth at small times. Note that our system is sensitive to DC signals since $F(\omega)$ is finite for $\omega\rightarrow0$.  For the purposes of this paper this DC
sensitivity is unwanted, and we will discuss methods of dealing with
it later.

It is important to note that we have waited until the coherent states
overlap fully. A measurement at a different time would suppress our signal by a factor of $A=e^{-\left(2\frac{r_0}{l_o} h_\bot(t)\right)^2}$.
The double-exponential suppression thus necessitates we obtain as complete
an overlap as possible. 
We can eliminate DC signals while simultaneously improving the overlap
and eliminating other sources of error through application of
$R_{\hat{y}}(\pi)$ pulses in a sequence analogous to the Carr-Purcell pulse
sequence in NMR~\cite{CarrPurcell}. If applied at a time when the velocity vanishes, a $\pi$ pulse
will time reverse the particle's motion, causing it to retrace its path. Such a
time is guaranteed and occurs at time intervals of $t = \frac{2
  \pi n}{\omega_+ + \omega_-} = \frac{\pi n}{\tilde{\omega}}$. If we
wait for an additional time time $2 t$ after the first $\pi$
pulse to apply a second $\pi$ pulse, the path will be time reversed
again and return to the origin. It is clear that any DC signals will
be canceled by such a pulse sequence as the path returns to itself so
the average position is zero. We note that this pulse sequence will also help to cancel certain noise sources such as Zeeman fields or small trapping asymmetries.

In the operator language such a pulse sequence has the form
\begin{equation}
U_{CP} = R_{\hat{y}}(-\pi/2) U(t) R_{\hat{y}}(\pi) U(2 t) R_{\hat{y}}(\pi) U(t) D[R\rr_0] R_{\hat{y}}(\pi/2). \label{eq:Ucp}
\end{equation}
 Application of such a pulse sequence modifies the response function to
\begin{equation}
F(\omega) = 2 i \sin(\omega t) \left[ F_0(\omega) e^{ i \omega t} + F_0^*(\omega) e^{-i \omega t} \right] e^{ i 2 \omega t}
\end{equation}
where $F_0(\omega)$ is the response function given in
Eq.~\ref{eq:f0}. The complex conjugate term $F^*_0(\omega)$ arises due
to the time reversal of the paths. This response function is plotted
in Fig.~\ref{fig:response}. The new response function now vanishes at $\omega=0$ and $\omega = \omega_\pm$, however we still have large sensitivity in the frequency range $\omega = \frac{1}{8} \frac{2\pi}{t}$ around $\omega_\pm$. 



We now generalize from a single particle to a thermal ensemble of ultracold
dilute atoms in the presence of an induced spin-orbit Hamiltonian of
the form in Eq. \ref{eq:H}. From a practical standpoint a system of dilute cold atoms allows us to
neglect complications arising atom-atom interactions in a BEC. 
Consider a thermal ensemble of cold atoms
at a temperature T. Using the Glauber P-representation~\cite{Glauber} the density
matrix has the form $\rho = \int \dd{\al} e^{- |\alpha_+|^2 / \ev{n_+}
  - |\alpha_-|^2 / \ev{n_-}} \ket{\al}\bra{\al}$, where $\ev{n_i} =
[\exp[\hbar \omega_i / k T] - 1]^{-1}$ is the average occupation for
the classical mode of frequency $\omega_i$. Such an ensemble
suppresses the expectation value of $S_z$ operator by a factor $e^{-
  \ev{n_+} |\gamma_+|^2 - \ev{n_-} |\gamma_-|^2}$ relative to the
single particle/zero temperature expectation value. The suppression
factor $\gamma_\pm$ depends on the pulse sequence used. For example,
using the pulse sequence $U_p$ given above we get $\gamma_+ =
\frac{l_o}{2}\int_0^{t} \dd{t^\prime} (g_x + i g_y) e^{i
  \omega_+ t^\prime}$ and $\gamma_- = \frac{l_o}{2}\int_0^{t} \dd{t^\prime}
(g_y + i g_x) e^{ i \omega_- t^\prime}$. Note that we can express $\int \dd{t} \rr \cdot \g$ as a superposition of $\gamma_+$ and $\gamma_-$.
 For the Carr-Purcell
like pulse sequence the suppression factor $\gamma_\pm$ is more
complicated, but can still forms a basis for which we can express the phase. This implies that $\gamma_\pm$ has a similar frequency dependence to $F(\omega)$. 

We are now in a position to discuss the measurement capabilities of such a system. We first estimate the maximum AC signal such a system can measure. To avoid signal suppression due to the finite temperature of the ensemble we bound the maximum strength of the AC signals our system can measure by $g_{max} \leq \frac{2 \gamma_{d}}{\sqrt{\ev{n}}} \frac{2 \pi \hbar}{ m r_0 }$. In this limit the sensitivity for our detector can be estimated with 
\begin{equation}
S \sim \sqrt{\frac{1}{N \tau}} \frac{2 \pi \hbar}{m r_0},
\end{equation}
where the lifetime of one measurement, $1/\tau = \gamma_{se} + \gamma_{coll}$ is limited by spontaneous emission, $\gamma_{se}$ and collisions, $\gamma_{coll}$. 

To minimize the sensitivity of the system we need to consider the effect of collisions and the spatial configuration of the system. We desire to confine our system to a ``laser homogeneity'' radius, $r_l$, for which nonlinear variations in the laser fields are suppressed. The effective $2D$ system will use an axial trapping potential of $\omega_\| \geq \log(2) kT / \hbar$ to freeze all motion into a single transverse mode. Thus our system will have $N_{l} = r_l / d$ layers, where $d = \sqrt{\hbar / m \omega_\|}$, providing for an increase in sensitivit of $1/\sqrt{N_{l}}$. The radius $r_l$ further constrains our sensitivity by bounding the maximum trap displacement by $r_0 = r_l - r_t$, where $r_t = \sqrt{\ev{n}} l_o \sim {\ev{v}}/{\tilde{\omega}}$ is the thermal radius of a thermal ensemble and $\ev{n} = kT / \hbar \omega $ and $\ev{v} = \sqrt{3 kT / m}$ are the respective high temperature thermal occupation number and velocity. (Fig.~\ref{fig:experiment})

The lifetime of the system will be dominated by spontaneous emission at low densities and collisions at high densities. To optimize the sensitivity we desire to place as many atoms per layer as possible. The collisional scattering rate is given by $\gamma_{coll} = \frac{N_a \ev{v} a^2} {d r_t^2}$, where $a$ is the interparticle scattering length. The critical number of atoms at which the collision rate begins to dominate the spontaneous emission rate is $N_c = \frac{\ev{v} a^2} {\gamma_{se} d r_t^2}$. We see that in the small atom number limit the sensitivity is a monotonically decreasing function of the trapping frequency and has a $1/\sqrt{N_a}$ dependence. However, in the large atom limit the sensitivity is minimized at a trapping frequency of  $\omega_{min} = 2 \ev{v} / r_l$ and the sensitivity becomes independent of the number of atoms per layer. Note that in this limit the bandwidth of the system is increased by adding atoms. (See Fig. \ref{fig:bw}.)

We assume our thermal ensemble has a temperature of $T \sim 1 \textrm{mK}$ with a frequency scale $\tilde{\omega} = 2 \pi \textrm{kHz}$. At these temperatures the gas is non-degenerate and is described well by a classical gas. For this temperature we find an upper bound of ${g} \sim 10^{-2} \textrm{m}/\textrm{s}^2$ before exponential suppression of the signal above becomes relevant. We will consider a cold gas of $^{87}\textrm{Rb}$ cooled to $T = 1 \mu\textrm{K}$ with an axial confinement distance of $d = 1 \mu\textrm{m}$. We take the spontaneous emission rate to be $\Gamma_{se} = 1 / 70 \textrm{ms}$~\cite{SMFSpielman1} and laser inhomogeneity radius to be $r_l = 10 - 25 \mu\textrm{m}$. In the $N_a \gg N_c$ limit we estimate the sensitivity to be $S \sim 10^{-7} \frac{\textrm{m}/\textrm{s}^2}{\sqrt{\textrm{Hz}}}.$
A similar analysis for a $3D$ system gives a sensitivity drop of approximately an order of magnitude. We note that had we instead used a fermionic species we would obtain a similar result since we have two spin species.


The concept of a continuous coupling of spin to momentum can also be extended to a continuous coupling of spin and position. We note that in a harmonic trap position and momentum are dual variables, and thus a spin-dependent term in the Hamiltonian that has spatial variation will experience a similar phase accumulation to the system described above. An example would be trapped spin-1 system in the presence of a Zeeman field with a strong spatial variation. In such a system the Zeeman field will act to trap (anti-trap) the $S_z=+1$ and $S_z = -1$ spin states with different trapping potentials. This will play a similar role to the opposite charge couplings to gauge fields given above. However, such a system requires strong magnetic field gradients so it may be impractical.


Finally, we note that this system is not limited to measurements of AC
signals. Through appropriate modifications of pulse sequences such a
scheme is capable of measurements of both gravitational gradients as
well as constant DC gravity. For example, the system will be sensitive
to DC signals if a measurement is made for any path at which the
average position perpendicular to the axis of displacement is
non-zero. Since overlap is guaranteed at the anti-node of the orbit,
this would be a reasonable measurement point. Similarly, the system is
sensitive to gravitational gradients if a measurement is made at a
point of overlap when no $\pi$ pulses have been applied. Both of these
effects will be eliminated through a Carr-Purcell like pulse
sequence. Due to electronics noise, the sensitivity of our system to
either DC or gradiometric signals will be significantly lower than
existing setups. Thus, we have chosen to not pursue them in detail.



{\em Acknowledgements:}~This research was supported by the JQI Physics Frontiers Center. BMA and VMG acknowledge support from US-ARO and JMT was supported through ARO-MURI-W911NF0910406. The authors are grateful to Ian Spielman for illuminating discussions.

\bibliography{acinterferometry}

\end{document}